\documentclass{moriond}

\def\be{\begin{equation}}
\def\ee{\end{equation}}
\def\bea{\begin{eqnarray}}
\def\eea{\end{eqnarray}}

\newcommand{\Photo}{}

\begin{document}
\vspace*{4cm}
\title{Search for resonances in four top quark events in the 2 lepton final state}

\author{ D.W. Stafford, on behalf of the CMS collaboration\footnote{Copyright 2026 CERN for the benefit of the CMS Collaboration. Reproduction of this article or parts of it is allowed as specified in the CC-BY-4.0 license.}}

\address{CMS Group, Deutsches Elektronen-Synchrotron DESY, Notkestr. 85, 22607 Hamburg, Germany}

\maketitle\abstracts{
A first search is presented for BSM resonances in four top quark production in the 2 lepton channel, using $138\mathrm{fb}^{-1}$ $pp$ data collected at $\sqrt{s}=13$~TeV, and $35\mathrm{fb}^{-1}$ $\sqrt{s}=13.6$~TeV $pp$ data. No significant excess is observed; limits are set on vector Z', scalar, pseudoscalar and ALP mediators. Z' mediators with 50\% width are excluded up to 850~GeV (1000~GeV expected).
}

\section{Motivation}

The observation of the four top quark process was announced by the ATLAS and CMS collaborations at the 2023 Moriond conference~\cite{ATLAS:2023ajo,CMS:2023ftu}. The cross section measured in these results, as well as previous measurements~\cite{CMS:2023zdh,ATLAS:2021kqb}, was higher than the predicted standard model (SM) cross section. This leaves space for additional beyond-the-SM (BSM) processes to also contribute to this final state. Furthermore, many BSM theories predict the existence of new mediators which couple predominantly or exclusively to top quarks, and hence would result in four top quark diagrams. Examples are top-philic Z' vector bosons~\cite{Greiner:2014qna}, scalar and pseudoscalar bosons in two Higgs doublet models~\cite{Branco:2011iw}, and Axion-like particles (ALPs)~\cite{Esser:2023fdo}.

\section{Analysis Strategy}

Previous searches have been performed for resonances in four top quark events in the 1-lepton channel using the LHC run 2 dataset (proton-proton ($pp$) collision data collected between 2015 and 2018)~\cite{ATLAS:2023taw,CMS-PAS-B2G-24-009}. The search reported here \cite{CMS-PAS-B2G-25-005} is performed using the full CMS run 2 dataset at a centre-of-mass energy $\sqrt{s}=13$~TeV, as well as the $pp$ data from 2022 at $\sqrt{s}=13.6$~TeV. While the luminosity of the 2022 dataset is relatively small ($35\mathrm{fb}^{-1}$, compared to $138\mathrm{fb}^{-1}$ for the run 2 dataset), the signal cross sections can be up to 30\% higher at $\sqrt{s}=13.6$~TeV than $\sqrt{s}=13$~TeV.

\begin{figure*}[htb]
\centering
\includegraphics[width=0.29\textwidth]{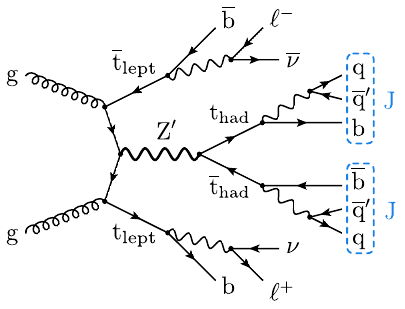}
\caption{\label{fig:selection-signature} The targeted $t\bar{t}Z'$ signal signature. It comprises two leptonically decaying top quarks and two hadronically decaying top quarks, reconstructed using a variable-radius jet clustering algorithm, denoted J.}
\end{figure*}

This search also targets a new topology, requiring the leptonic decay of both top quarks produced in association with the new mediator, and the hadronic decay of the top quarks produced by the mediator (Fig. \ref{fig:selection-signature}). The selection of two leptons helps suppress many common backgrounds from quantum chromodynamics (QCD), while the hadronic decay of the top quarks from the mediator means this can be reconstructed entirely from visible decay products.

To select the leptonic decays of the top quarks, two leptons are required, as well as two small radius jets. At least one of these jets is required to be tagged as coming from the a b quark (b-tagged), and the signal regions are split into categories for one and at least two b-tagged jets.

The top quarks from the resonance decay are each reconstructed as large radius jets capturing all hadronic decay products of one top quark. The most commonly used large radius jets in CMS, anti-$k_{T}$ jets with a radius $R=0.8$, are only efficient at capturing very high $p_{T}$ top quarks, achieving 80\% efficiency for $p_{T}>800$~GeV, which is much higher than the $p_{T}$ of many top quarks jets in this signature. However, using a larger jet radius could cause other objects in the event to be vetoed due to overlap. Therefore HOTVR jets~\cite{Lapsien:2016zor} are used, which have a radius which scales proportionally to the inverse of the $p_{T}$.

A new boosted decision tree (BDT) based top tagger was developed for this search~\cite{CMS-DP-2024-038}, which provided significantly improved signal performance compared to the previously available cut-based tagger - for instance 32\% compared to 20 \% for the low-$p_{T}$ ($200<p_{T}<400$~GeV) working point. A further advantage of the new tagger is a new working point could be selected - it was found that a working point with 84\% signal efficiency and a mistagging rate of 22\% was most effective due to the very low signal rates in this search.

After these cuts, the main background in the signal regions (SRs) is top quark pair production ($t\bar{t}$) events, where the tops decay in the leptonic channel, and additional jets are mistagged as top HOTVR jets. This background was estimated by applying a transfer factor, derived from simulation, to events in regions with an identical selection to the SRs, except requiring 0 HOTVR jets to be top tagged. There is also a small background contribution from events where at least one of the HOTVR jets is initiated from a top quark, which is modelled directly from simulation.

\begin{figure*}[htb]
	\centering
	\includegraphics[width=0.58\textwidth]{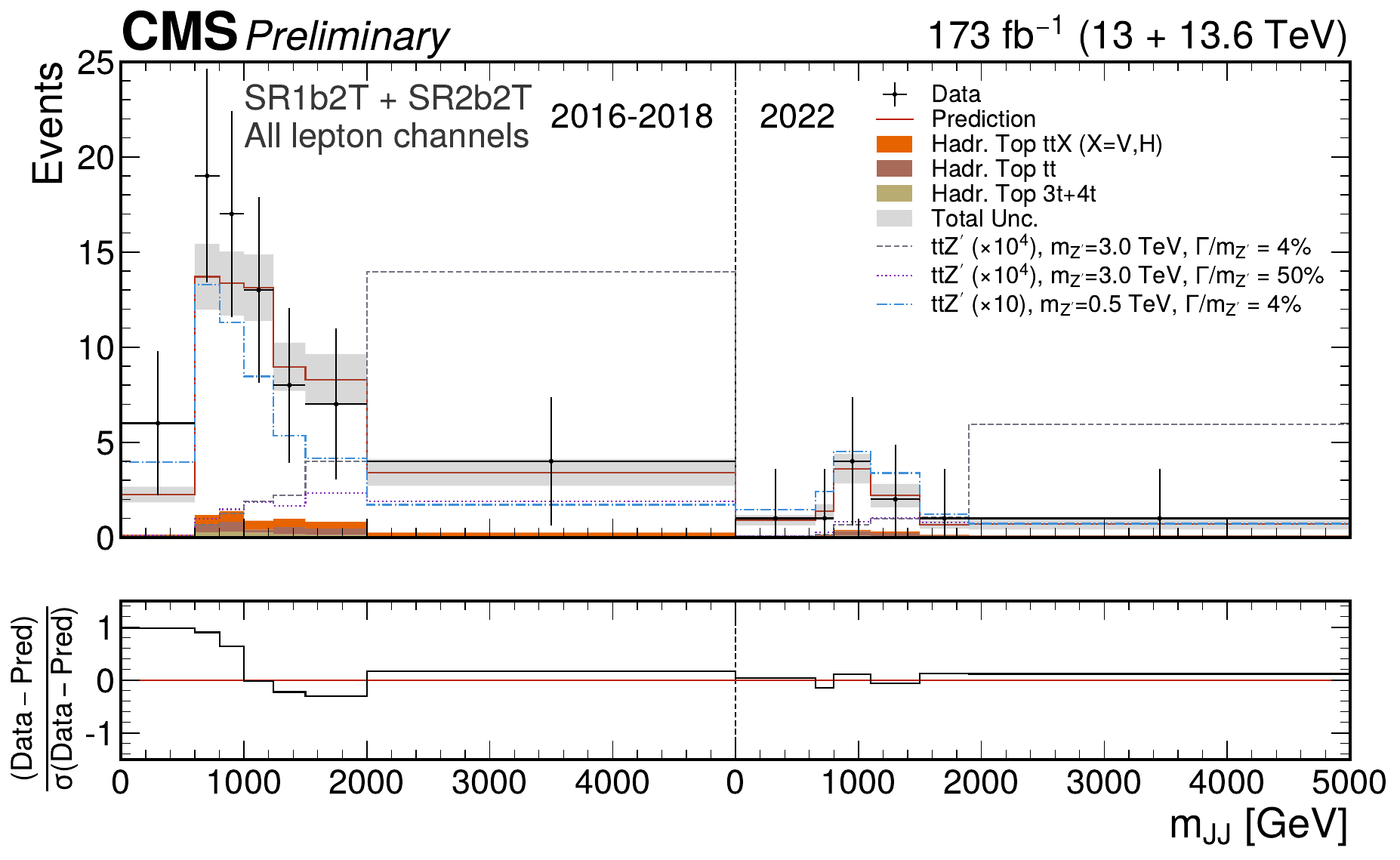}
	\caption{\label{fig:postfitdistributions}The background-only post-fit distribution of $m_{JJ}$ for the background prediction (red line), including contributions estimated directly from simulation (filled bins) in the merged signal regions for the 2016--2018 (left) and 2022 (right) data-taking periods.}
\end{figure*}

A fit is then performed on the invariant mass of the two leading HOTVR jets, $m_{JJ}$, split in channels with different lepton flavours, b jet categories and centre of mass energies. The fit distribution, combined across lepton flavours and b jet categories, is shown in Fig. \ref{fig:postfitdistributions}.

\section{Results}

\begin{figure*}[htb]
	\centering
	\includegraphics[width=0.3\textwidth]{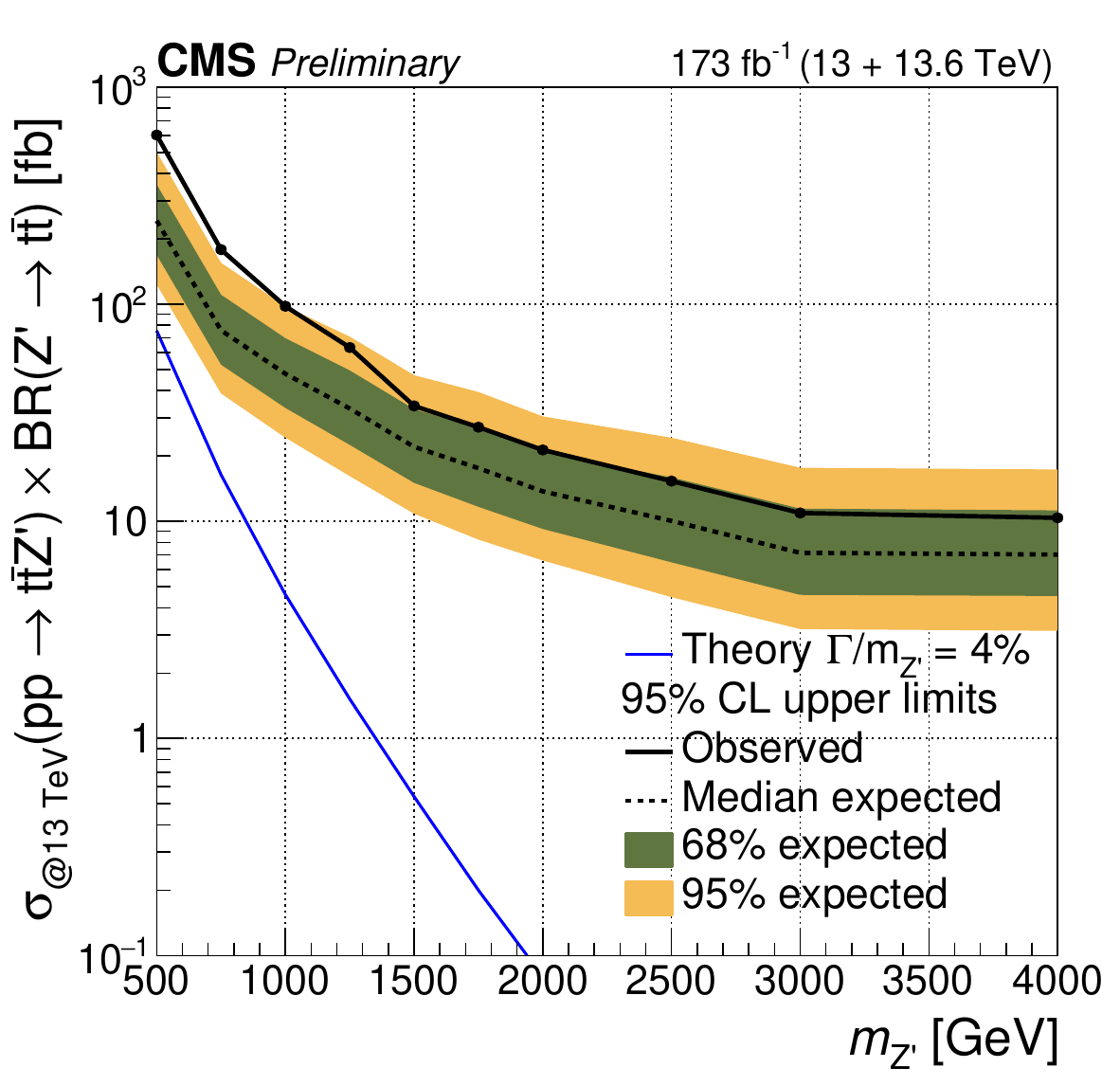}\hspace{0.08\textwidth}
	\includegraphics[width=0.3\textwidth]{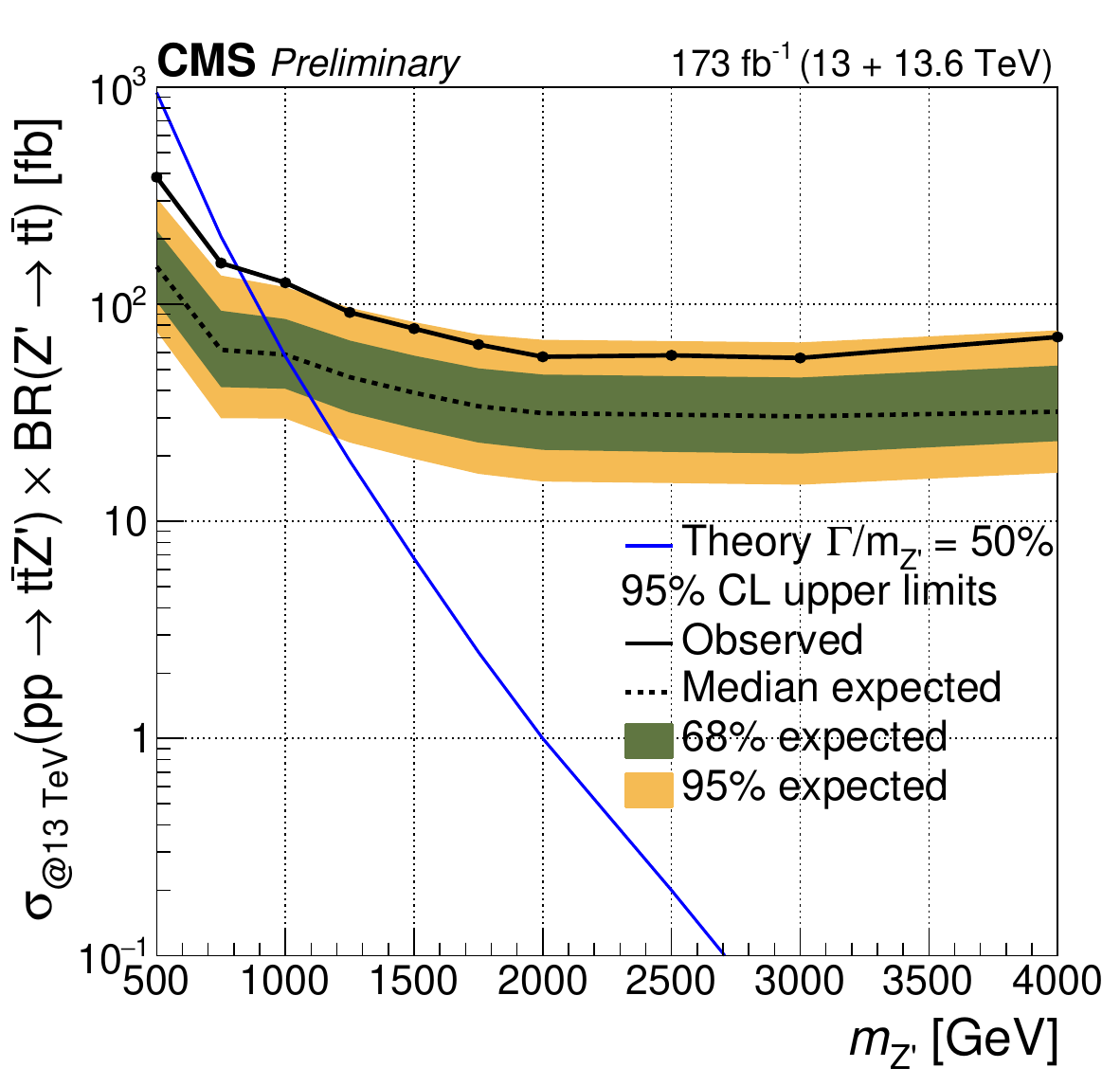}
	\caption{\label{fig:result-limitzprime}Expected and observed upper limits at 95\% confidence level on the $t\bar{t}Z'$ production cross section times branching fraction at 13~TeV as a function of $m_{Z'}$ for 4\% (left) and 50\% (right) relative decay widths.}
\end{figure*}

No significant excess was observed, and limits were therefore set on the $t\bar{t}Z'$ process, as show in Fig. \ref{fig:result-limitzprime}. Since there is an upwards fluctuation of the data in the first bins of the fit distribution, the limits are weaker than expected, particularly for the lower mediator masses, with a largest significance of 2.2 standard deviations for the 500~GeV, 4\% width mediator.

\begin{figure*}[htb]
	\centering
	\includegraphics[width=0.3\textwidth]{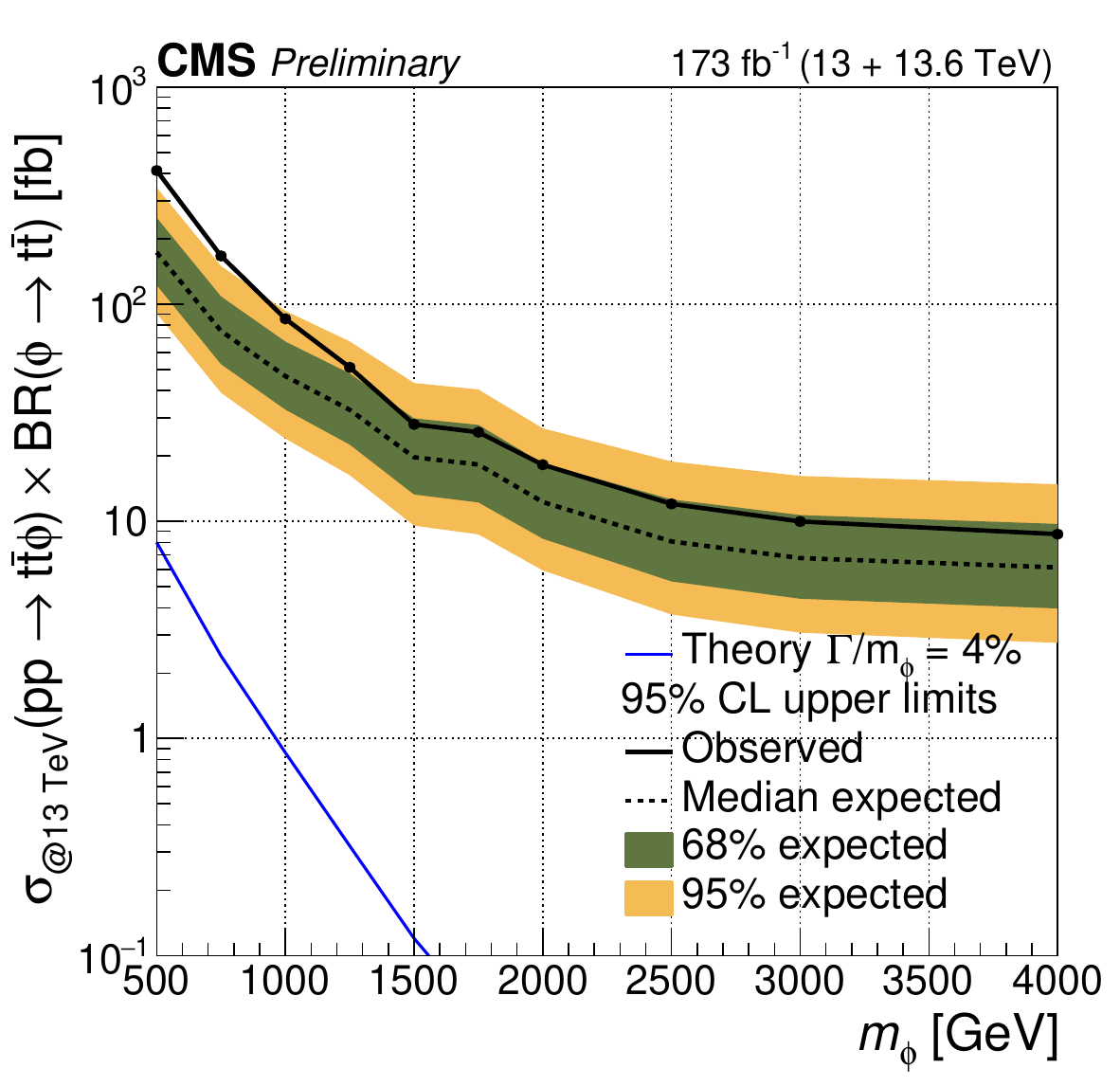}\hspace{0.08\textwidth}
	\includegraphics[width=0.3\textwidth]{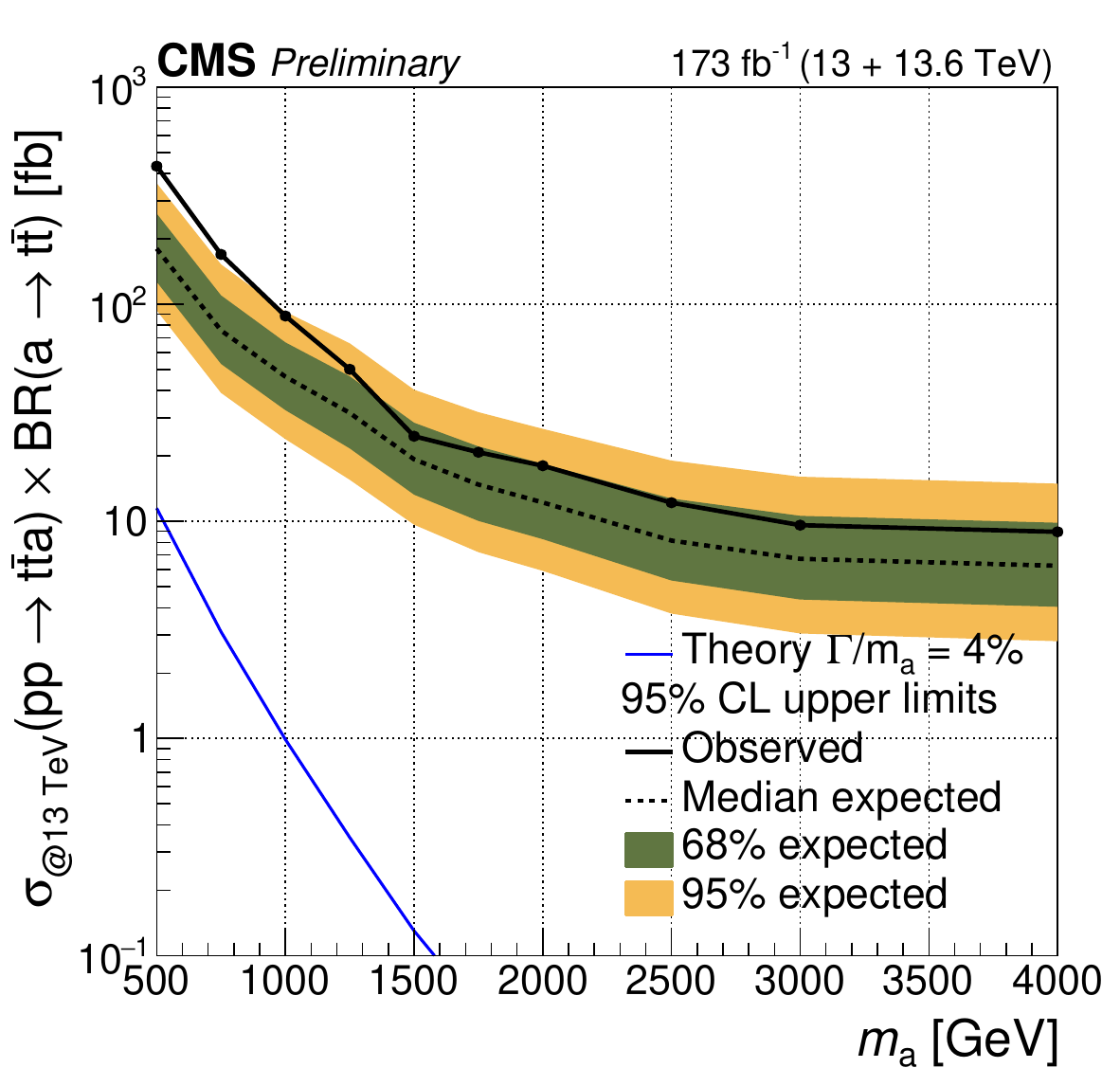}
	\caption{\label{fig:result-limitttah}Expected and observed upper limits at 95\% confidence level on the $t\bar{t}\phi$ (left) and $t\bar{t}a$ (right) production cross section times branching fraction at 13~TeV as a function of mediator mass for the 4\% relative decay width. }
\end{figure*}

Limits are also set on the scalar and pseudoscalar processes, $t\bar{t}\phi$ and $t\bar{t}a$, as shown in Fig. \ref{fig:result-limitttah}. These are very similar to the limits on $t\bar{t}Z'$, as the kinematic properties only differ slightly.

\begin{figure*}[htb]
	\centering
	\includegraphics[width=0.45\textwidth]{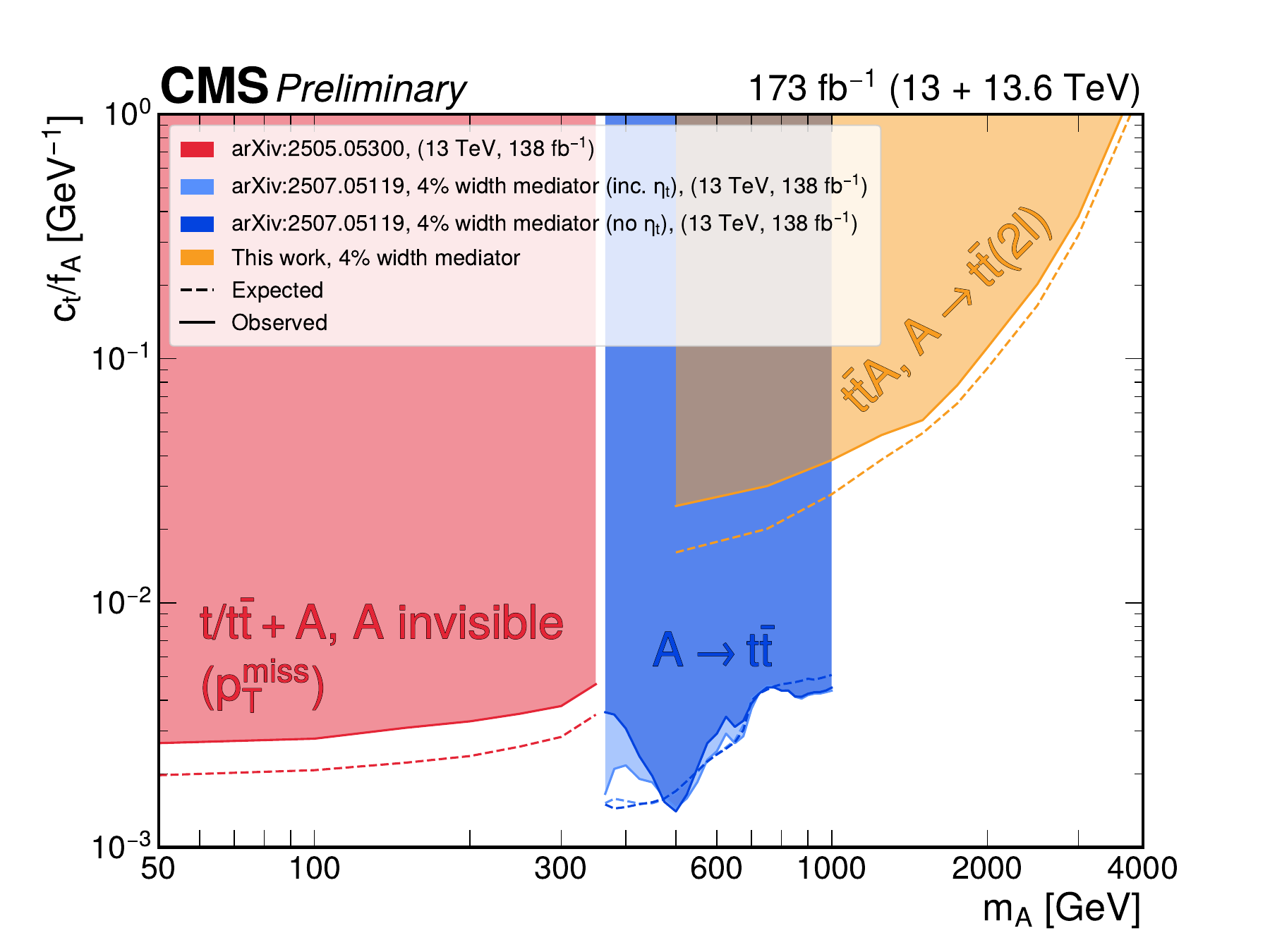}
	\caption{Expected (dashed lines) and observed (solid lines, shaded areas) upper limits at 95\% confidence level on ALP-top coupling divided by the decay constant, $c_t/f_{A}$, from: red - t/$t\bar{t}+p_{T}^{\mathrm{miss}}$ limits~\protect\cite{CMS:2025ncs}; light blue - $t\bar{t}$ resonance limits~\protect\cite{CMS:2025dzq} including a contribution from non relativistic QCD, $\eta_{t}$, in the background and, darker blue, the limits when not considering $\eta_{t}$ in the background; orange - four top quark limits from the search described here.}
    \label{fig:result-limitALPs}
\end{figure*}

Finally, limits are set assuming the pseudoscalar mediator is an axion like particle. In the spirit of the axion limits project, these limits are compared with two other CMS results, as shown in Fig. \ref{fig:result-limitALPs}. The first are limits from a search for pseudoscalar (and scalar) resonances in $t\bar{t}$ production~\cite{CMS:2025dzq}, and the second is a search for $t\bar{t}$ or single tops produced in association with missing transverse momentum~\cite{CMS:2025ncs} - assuming the ALP only couples to top quarks, it would have a very long lifetime below the top quark threshold, and would hence tend not to be detected within the CMS experiment.

\section{Conclusion}

A first search has been presented for BSM resonances in four top quark production in the 2 lepton channel, which is also the first four top quark search to include data collected at $\sqrt{s}=13.6$~TeV. A novel BDT top tagger for variable-radius jets is used to improve signal acceptance. No significant excess is observed, and so limits are set - the 50\% width Z' mediator is excluded up to 850~GeV (1000~GeV expected). Limits are also set on scalar, pseudoscalar and ALP mediators. The analysis is statistically limited, so improvements are expected in future analyses using the full CMS run 3 dataset.

\section*{Acknowledgements}

DWS acknowledges support from DESY, a member of the Helmholtz Association HGF.

\section*{References}
\bibliography{moriond}

\end{document}